\documentclass[11pt]{article}
\usepackage[top=1in, bottom=1in, left=1in, right=1in]{geometry}

\usepackage{times}
\usepackage{soul}
\usepackage{url}
\usepackage[utf8]{inputenc}
\usepackage{amsmath}
\usepackage{amssymb}
\usepackage{amsthm}
\usepackage{booktabs}
\usepackage{algorithm}
\usepackage[noend]{algorithmic}
\usepackage{xcolor}
\usepackage{dsfont}
\usepackage{hyperref}
\usepackage[round]{natbib}
\usepackage{parskip}
\usepackage{mathpazo}

\usepackage{graphicx}

\usepackage{url}

\usepackage{xr}

\newcommand{\E}{\mathbb{E}}
\newcommand{\R}{\mathbb{R}}
\newcommand{\X}{\mathbf{X}}
\newcommand{\x}{\mathbf{x}}
\newcommand{\w}{\mathbf{w}}
\newcommand{\F}{\mathcal{F}}
\newcommand{\V}{\mathbb{V}}
\newcommand{\Z}{\mathbb{Z}}
\renewcommand{\H}{\mathcal{H}}

\newcommand{\set}[1]{\{#1\}}

\newcommand{\norm}[1]{\left| |#1|\right|}
\newcommand{\ip}[1]{\langle #1 \rangle}
\newcommand{\citea}[1]{\citeauthor{#1} (\citeyear{#1})}
\newcommand{\vk}[1]{\textcolor{black}{#1}}

\usepackage{color-edits}
\addauthor[Hoda]{hh}{red}
\addauthor[Vince]{vc}{blue}

\title{On the Pros and Cons of Active Learning for Moral Preference Elicitation}

\usepackage{authblk}

\author[1]{Vijay Keswani}
\author[2,3]{Vincent Conitzer$^*$}
\author[2]{Hoda Heidari$^*$}
\author[1]{Jana Schaich Borg$^*$}
\author[1]{Walter Sinnott-Armstrong$^*$}

\affil[1]{Duke University}
\affil[2]{Carnegie Mellon University}
\affil[3]{University of Oxford}

\date{}

\begin{document}

\maketitle
\def\thefootnote{*}\footnotetext{These authors contributed equally to this work and are ordered alphabetically.}\def\thefootnote{\arabic{footnote}}

\begin{abstract}
Computational preference elicitation methods are tools used to learn people's preferences quantitatively in a given context. Recent works on preference elicitation advocate for \textit{active learning} {as an efficient method} to iteratively construct queries (framed as comparisons between context-specific cases) that are likely to be \textit{most informative} about an agent's underlying preferences. In this work, we argue that the use of active learning for moral preference elicitation relies on certain assumptions about the underlying moral preferences, which can be violated in practice. Specifically, we highlight the following common assumptions (a) preferences are stable over time and not sensitive to the sequence of presented queries, (b) the appropriate hypothesis class is chosen to model moral preferences, and (c) noise in the agent's responses is limited. While these assumptions can be appropriate for {preference elicitation in certain domains}, prior research on moral psychology suggests they may not be valid for moral judgments.
Through a synthetic simulation of preferences that violate the above assumptions, we observe that active learning can have similar or worse performance than {a basic random query selection method} in certain settings. Yet, simulation results also demonstrate that active learning can still be viable if the degree of instability or noise is relatively small and when the agent's preferences can be approximately represented with the hypothesis class used for learning. Our study highlights the nuances associated with effective moral preference elicitation in practice and advocates for the cautious use of active learning as a methodology to learn moral preferences.

\end{abstract}

\section{Introduction}
Ensuring proper deployment of artificial intelligence (AI) systems in high-stakes societal domains requires building trust in the decisions of these systems.
To that end, recent work on ethical and participatory algorithmic development emphasizes the importance of encoding stakeholders' {values} in these systems, especially their {moral {judgments/}preferences} over actions that can cause significant harm to others \cite{feffer2023preference}.
Incorporating stakeholders' {moral preferences} allows for the creation of tools whose judgments are normatively aligned with those of the stakeholders and helps counter various harms associated with the use of computational tools.
To accomplish this goal, however, one first needs to accurately elicit people's moral preferences.

Studies on moral preference elicitation
often present agents
with pairs of context-specific cases and ask them to choose the
one they prefer. 
Using the agent's responses for a set of such \textit{pairwise comparisons}, one can try to learn a representation of their underlying preferences.
To formalize this setting, let $\X \subseteq \R^d$ denote the space of all {cases} over which any agent has preferences, with $d \in \Z_{+}$ denoting the number of features describing each {case}.
Following the standard preference elicitation literature, 
suppose that an agent's preferences are {determined by comparing the value of} an underlying utility function $u : \X \rightarrow \R$ {across cases} \cite{freedman2020adapting}. 
For any input pair $\x, \x' \in \X \times \X$, the agent prefers $\x$ over $\x'$ iff $u(\x) > u(\x')$. 
Let $R: \X \times \X \rightarrow \set{0,1}$ denote their response function, 
with $R(\x, \x') = \mathds{1}(u(\x) > u(\x'))$, where $\mathds{1}(\cdot)$ is the indicator function.

\begin{figure}
    \centering
    \fbox{\includegraphics[width=0.5\linewidth]{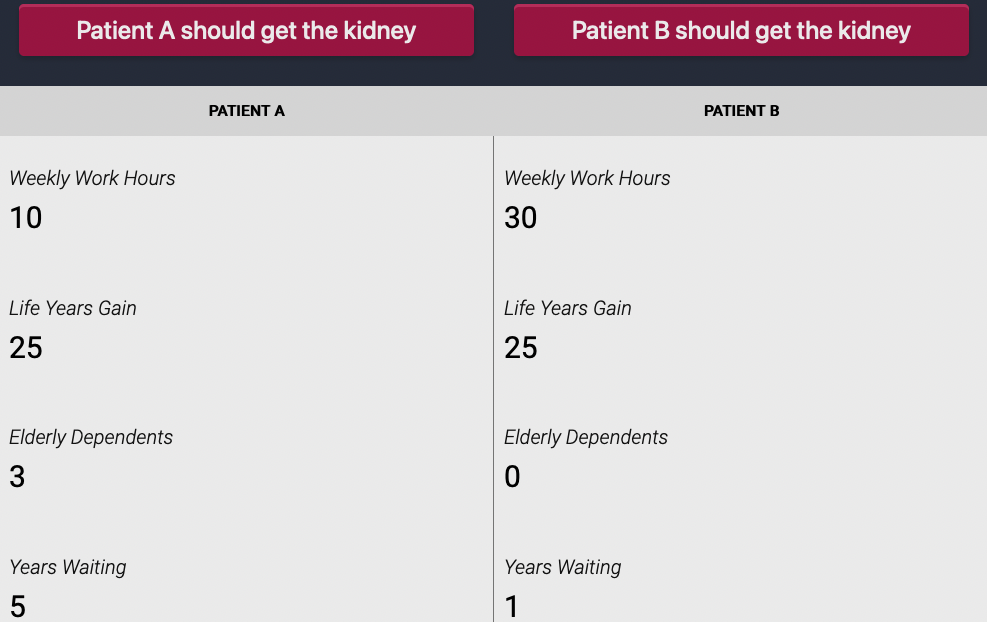}}
    \caption{Example of a pairwise comparison from the \citet{boerstler2024instability} study on kidney allocation decisions.}
    \label{fig:kidney_example}
\end{figure}

Multiple recent studies employ this framework for moral preference elicitation.
For example, {\citet{boerstler2024instability}
model lay-agent's moral preferences in kidney allocation. 
They provide participants with profiles of two patients who need kidney transplants and ask them to decide which patient should receive the one available kidney.
Each patient profile contains features like the patient’s number of children, years of life they will gain from the transplant, etc. The choice between the two patients can pose a moral dilemma when different features favor different patients \cite{sinnott2021ai} (Figure~\ref{fig:kidney_example} presents a pairwise comparison scenario from this study).
Another well-known example of this approach is the ``Moral Machines'' study,
in which participants are presented with sacrificial moral dilemmas and asked what an autonomous vehicle should do in each case \cite{awad2018moral, noothigattu2018voting}. In another study, 
 % similarly 
\citea{srivastava2019mathematical} elicit \textit{fairness preferences} by presenting participants with pairwise comparisons of algorithmic predictions and backing out the notion of fairness that is most compatible with their responses.
Preference elicitation has similarly been part of the development pipeline of various \textit{participatory} computational frameworks \cite{lee2019webuildai,kahng2019statistical,loreggia2019metric,feffer2023preference}.
}

The goal of preference elicitation in these settings is to 
accurately and efficiently 
learn a
representation of the agent's underlying utility $u(\cdot)$ using their responses for a given set of $N$ pairwise comparisons, i.e., using $\set{(\x_t, \x_t', R(\x_t, \x_t'))}_{t=1}^N$. 
Here, accuracy refers to the ability to {(a) recover the utility function $u$ 
and/or (b) offer an approximate representation of $u$ that mimics decisions made through $u$ in a large number of comparisons.}
Achieving accuracy often requires presenting an agent with numerous pairwise comparisons, which can be onerous and expensive. 
To reduce the number of queries required to obtain a desired level of accuracy, \textit{active learning} is {frequently} invoked as an alternative approach. 

Active learning methods operate in the realm of scarce outcome-labeled data, where one has the option to interactively query an \textit{oracle} (the user/agent in this case) for labels, and the goal 
is to learn the relationship between labels and relevant features using as few queries as possible \cite{settles2009active}.
These methods can help improve the efficiency of preference elicitation as well.
For preference elicitation, active learning techniques can 
suggest new pairwise comparisons that would provide the \textit{maximal information} about the agent's utility function {given the information gathered so far} \cite{dragone2018constructive}.
Using this form of structured determination of the next pairwise comparison (based on the agent's previous responses), the agent's preferences can be inferred faster than the setting where they are presented with random comparisons at each time step.
For this reason, multiple recent works consider active-learning-based preference elicitation. \citea{yang2021bayesian} use interactive elicitation to create recommendation systems.
\citea{srivastava2019mathematical} develop active-learning-based surveys to elicit fairness preferences.
\citea{johnston2023deploying} use active learning to learn preferences regarding healthcare resource allocation.
These recent use cases of active learning provide evidence of its ability to efficiently elicit people's preferences. \textbf{However, the effectiveness of active learning 
relies on certain assumptions that may not hold in the case of moral preferences}.

Moral preferences capture a person's normative views over available actions in moral dilemmas---that is, what is the \emph{right} thing to do when the chosen action could lead to significant harm to others, but not (or not only) to the participant themself? A popular example is the \textit{trolley problem}, where the participant is asked which human lives should be prioritized, passengers or pedestrians \cite{foot1967problem}.
{
Similarly, in the kidney allocation example described earlier, when asked to decide which of two patients on the kidney transplant list should get the kidney, a participant's decisions are based on patient features that they consider \textit{morally relevant}.
In these settings, when an agent expresses a preference for one patient over another, their judgment can be characterized by the underlying \textit{utility function} they use to assign \textit{relevance scores} to the available actions, choosing the action with the highest assigned score.
\vk{Note that, despite the use of utility functions, this standard setup does not presuppose any utilitarian moral theory because it can model agents who base their decisions on non-utilitarian factors, such as past misbehaviors by patients.}
Modeling the participant's preferences in moral decision-making settings (e.g., by learning their underlying utility function) allows for predicting their moral judgments when presented with new dilemmas in the same setting. Therefore, these models can be useful in the development of ethical AI tools \cite{feffer2023preference}.\footnote{A note on terminology: what we call \emph{moral preferences} can also be described as judgments/orderings over available actions in moral dilemmas. 
This characterization is different than decision theory literature, which defines preference orderings over outcomes rather than actions \cite{arrow1996rational}.
Yet, we use the term preferences to be consistent with CS preference elicitation literature on modeling decision processes in pairwise comparison settings.
}
However, eliciting moral preferences can be challenging, and differ from the process of eliciting other kinds of preferences.
}

{
{Moral preferences} concern harms to others, and differ from
{self-interested, economic, or material} preferences, where the agent chooses the option with the highest subjective utility to self \cite{capraro2021mathematical}.
Instead of being concerned only with the self, moral preferences are intended to be impartial \cite{vanberg2008economics} and fair \cite{bicchieri2010behaving}.
Computational modeling of these preferences can, therefore, help develop decision-aid tools that incorporate stakeholders' moral values, e.g., in applications like autonomous vehicles or biomedical situations. 
Unsurprisingly, the standards of expected elicitation accuracy in these domains are quite high, since
inaccurate prediction of moral judgments can significantly harm the people using or affected by the decision-aid tool.
For these reasons, greater attention to elicitation performance is required in moral decision-making settings.

Yet, a crucial problem in eliciting moral preferences is that
they can be \textit{unstable}, i.e.,
{the participant's} choices for the first few presented moral dilemmas might appear inconsistent with each other
\cite{crockett2016computational}.
The participant can also be indecisive and provide ``noisy'' judgments to moral dilemmas (e.g., there may be variability in their choices for similar scenarios), further complicating the elicitation process
\cite{rehren2022stable}.
Research on moral psychology also lacks consensus on the structure of cognitive processes that incorporate moral preferences within our judgments \cite{ugazio2022neuro}. Limited understanding of moral decision-making structures makes it difficult to model them computationally.
All these properties taken together make moral preference elicitation a complex task and call into question the validity of active learning as a reliable elicitation methodology.
}

The use of active learning for preference elicitation often presupposes that the context in question does not suffer from the above issues.
Preference stability, limited variability in responses, and availability of a hypothesis class that captures the underlying utility $u$
are common assumptions \cite{dragone2018constructive}.
An obvious question that then arises is \textit{whether active learning still leads to efficient {moral} preference elicitation when these assumptions are violated.}
Research from moral psychology suggests that these assumptions may specifically not hold for moral preferences.
Hence, the efficacy of active learning for moral preference elicitation needs further examination.

\subsection{Our Contributions}
In this paper, we investigate whether active learning can be effective for moral preference elicitation, based on simulations designed to replicate the above challenges {with moral preferences}.
Our simulations test two popular active learning paradigms, version-space-based active learning and Bayesian active learning (Section~\ref{sec:model}).
Inspired by recent human subject research on properties of moral decision-making (Section~\ref{sec:challenges}), we consider the following challenges: (a) \textbf{preference instability}, (b) \textbf{model misspecification}, and (c) {\textbf{noisy responses}}.
In all settings, we compare active-learning-based approaches against a standard approach that presents agents with random pairwise comparisons. {We observe the following:}
    % \item 

    \paragraph{Preference instability.} Our simulations here evaluate elicitation performance when the agent's moral preference model stabilizes only after responding to a certain number of initial comparisons (Section~\ref{sec:pref_change}).
     Specific scenarios we consider include{: (1)} the agent, after a few comparisons, simplifies their moral preference to reduce the decision-making effort, {(2) the agent }makes their preference more complex to incorporate additional information, {and (3) the agent} changes their preference {entirely} to reflect {significant} updates to their moral values. 
    We observe that, in all three cases, when the number of features is small, the Bayesian active learning approach recovers well from instability and achieves higher accuracy 
    than the random query baseline within a small number of comparisons after a preference change.
    However, in cases of drastic preference changes and a large number of features, both active learning approaches have similar or worse performance than the random query baseline
    due to their dependence on previous comparisons. 
    \emph{The key takeaway here is that
    the accuracy and efficiency of active learning depend on the expected scale of preference instability (as captured by the kind of preference change) and the complexity of the decision-making context (as captured by the number of features).
    }   

    \paragraph{Model misspecification.} Our model misspecification simulation evaluates preference elicitation performance when the agent's moral decision-making model
and the model class used by the elicitation framework are different (Section~\ref{sec:model_misspecify}).
For instance, suppose the agent uses a shallow decision tree to encode preferences, but the preference elicitation framework uses the class of linear models.
Here, active learning {\em at best} converges to the best hypothesis in the linear class
but has a relatively high predictive error--as observed in our simulations.
Along with agents that use tree-based models, we simulate other scenarios of model misspecification, such as scenarios where the agent uses feature interactions but the elicitation model doesn't, and scenarios where the agent and the elicitation model use different feature sets. 
When the extent of model misspecification is large, we observe that active learning approaches and random query baseline have similar performance.
\emph{The key takeaway here is that appropriate modeling of the agent's moral decision-making process is necessary for active learning to
improve the elicitation efficiency of the framework. }

\paragraph{Noisy responses.} We also consider the setting where the agent's responses are stochastic and
simulate two kinds of stochasticity: (a) \textit{response noise}: when stochasticity in the agent's response to a pairwise comparison depends on the difference between utility assigned to each item in the pair (i.e., higher variability when utilities are close)
and (b) \textit{preference noise}: when the agent's preference model is sampled from a certain distribution.
For response noise, we observe that the Bayesian active learning approach is still more efficient than the random query baseline despite noise.
For preference noise, active learning is more efficient than random query baseline only when noise 
is \textit{small} (e.g., when noise magnitude is small relative to the range of model parameter values).
\emph{The key takeaway here is that one needs to consider the source and impact of variability in agent responses to assess the effectiveness of active-learning-based elicitation.}

\vspace{0.1in}
\noindent
Overall, our simulations shed light on the performance of active learning for simulated moral preference elicitation tasks.
We find that active learning can improve elicitation efficiency in certain settings (e.g., small-scale noise) but also reduce elicitation efficiency in other settings (e.g., large-scale preference instability).
Based on these results, we emphasize the need to understand the nuances associated with the moral decision-making context in question before deploying active learning-based elicitation frameworks. 
Additionally, our findings %through these simulations 
can inform future human-subject studies aimed at understanding the extent to which these assumptions are violated in common moral preference elicitation tasks.
% \footnote{\vk{The technical appendix sections for this paper are available in the extended version on Arxiv.
% -- https://arxiv.org/abs/.}
% }}

\subsection{Related Work}
Preference elicitation methods are employed in multiple domains to create user-centered services, e.g.
to create recommendation systems \cite{priyogi2019preference},  to understand consumer behaviour \cite{ben2019foundations}, and for {patient-centered} decision-making in healthcare \cite{weernink2014systematic}.
Research on preference elicitation similarly 
spans multiple disciplines,
including computer science \cite{chen2004survey}, economics \cite{beshears2008preferences}, and psychology \cite{slovic2020construction}.
Machine-aided elicitation 
has further improved learning efficiency by helping process available agent data and/or the choices they make in real and hypothetical scenarios \cite{soekhai2019methods}.
As mentioned earlier, similar efforts have been made in moral domains, with several applications employing elicitation frameworks to model moral preferences \cite{awad2018moral, srivastava2019mathematical, loreggia2019metric,balakrishnan2019using, sinnott2021ai,johnston2023deploying}.
For a general survey of {moral} preference elicitation methods, we recommend  
\citea{feffer2023preference}.
In our work, we focus on 
methods that query an agent to choose between two given cases 
and use their responses to learn their preferences \cite{ben2019foundations}. 
While pairwise comparisons are a popular elicitation technique, there are alternative approaches as well,
e.g., asking agents to report their preference strength \cite{toubia2003fast}, rank choices \cite{ali2012ordinal}, participate in bidding processes \cite{conen2001preference}, or describe the motivations for their choices \cite{liscio2023inferring,liscio2024value}.

Active learning can be used
to either learn the agent's utility model or to successively present them with better recommendations
\cite{houlsby2011bayesian,dragone2018constructive}.
For the former setting learning the utility model, \citea{huang2016consumer} propose active learning methods to learn marketplace consumer preferences and \citea{srivastava2019mathematical} elicit fairness preferences using active-learning-based surveys.
For the latter setting of generating personalized recommendations, \citea{elahi2014active} and \citea{yang2021bayesian} discuss active learning strategies to streamline data collection for recommendation systems.
\citea{johnston2023deploying} use uncertainty-based active learning methods proposed by \citea{vayanos2020robust} to model healthcare resource allocation preferences of survey participants.
Our work focuses on learning the utility model since the eventual goal is to use the learned utility and preferences for downstream applications.

Most preference elicitation studies focus on preferences involving self-benefits, e.g., to create recommendation systems or better-personalized services.
As mentioned earlier, moral preferences go beyond self-interest and explain people’s normative impartial judgments.
For instance, \citea{bicchieri2010behaving} show the insufficiency of monetary preferences in explaining people’s fairness perceptions.
\citea{capraro2018right} discuss how social preference models can be incompatible with people’s choices of equitable actions.
Other experimental analyses from psychology (see \citea{capraro2021mathematical} for a review) provide further evidence of contrasts between moral and material preferences.

Beyond our work, certain recent papers examine the limitations of active learning in different contexts.
\citea{margatina2023limitations} and \citea{kottke2019limitations} 
discuss the dependence of active learning's performance on common (but potentially unrealistic) assumptions, e.g. representative training data and equal labeling costs across cases.
Active learning can also fail to outperform random query baselines when faced with distribution shifts  \cite{snijders2023investigating} or outliers \cite{karamcheti2021mind}.
Data collected using active learning is implicitly tied to the learning model and can lead to generalization issues \cite{lowell2019practical}.  
Our work adds to this line of research, specifically questioning the applicability of active learning to moral preference elicitation.

\section{Algorithms for Preference Elicitation}
\label{sec:model}

The basic structure of the 
elicitation procedure is described in Algorithm~\ref{alg:pref_elicit}.
At time-step $t$, the agent is presented with a sampled comparison $(\x_t, \x_t')$ and their response 
is recorded. 
Then, the algorithm finds the hypothesis $h_t$ from class $\H$ which best fits the labeled comparisons recorded till time $t$.

\begin{algorithm}[tb]
\caption{Online preference elicitation}\label{alg:pref_elicit}
    \textbf{Input}: Functions $\text{sample}(\cdot)$, 
    $R(\cdot, \cdot)$, and $\text{fit}(\cdot, \cdot)$, $N$, class $\H$
    
\begin{algorithmic}[1]
\STATE $S \gets \emptyset$\\
\FOR{$t \in \set{1, \dots, N}$}
    \STATE $\x_t, \x_t' \gets \text{sample}(S)$  \COMMENT{sample new comparison} \\
    \STATE $r_t \gets R(\x_t, \x_t')$  \COMMENT{Get agent's response} \\
    \STATE $S \gets S \cup \set{(\x_t, \x_t',r_t)}$   \\
    \STATE $h_t \gets \text{fit}(S, \H)$  \COMMENT{Learn hypothesis using dataset $S$} \\
\ENDFOR 
\STATE \textbf{return} $h_N$
\end{algorithmic}
\end{algorithm}

The sampling step of Algorithm~\ref{alg:pref_elicit} (Step 3) can be executed either by randomly sampling a pair of input {cases} or by using \emph{active learning}, whereby the chosen pair depends on the comparisons presented so far and the hypothesis class $\H$.
We will use \textbf{\textsc{Random-PE}} to refer to the instance of Algorithm~\ref{alg:pref_elicit} that uses random sampling.
When using active learning to sample comparisons, multiple methods from prior works can be employed and we outline two popular approaches below.
Longer mathematical descriptions and use cases of these methods are provided in Appendix~\ref{sec:active_descriptions}.

\paragraph{Version-Space-based Active Learning.} 
Given a kernel-SVM decision boundary learned using available labeled data, the informativeness of any new query can be approximated using the distance of the query from the decision boundary and this heuristic can be used to generate an informative next query
\cite{tong2001support}.
To implement this approach,
we learn an SVM hypothesis $f$ that best fits $((\x_i, \x_i'))_{i=1}^t$ to labels $(r_i)_{i=1}^t$ and find a comparison that 
is closest to $f$'s decision boundary.
We will call this approach \textbf{\textsc{Active-VS-PE}}.

\paragraph{Bayesian Active Learning.} 
The Bayesian Active Learning with Disagreement (BALD) algorithm represents preferences using a Gaussian process with a specified kernel and chooses the next query to be the one that maximizes the mutual information between model predictions and model posterior \cite{houlsby2011bayesian}.
Implementation of this approach for Algorithm~\ref{alg:pref_elicit} requires learning a representation of the posterior corresponding to the labeled dataset $((\x_i, \x_i', r_i))_{i=1}^t$ and then finding the pairwise comparison with high mutual information.
We will call this approach \textbf{\textsc{Active-Bayes-PE}}.
Note that the use of learned posterior is limited to the sampling step and can be independent of the learning step.

% \vspace{0.05in}
The final step of Algorithm~\ref{alg:pref_elicit} (Step 6) uses a pre-specified function $\text{fit}(\cdot, \cdot)$ to learn a hypothesis $h_t$ from $\H$ that ``best'' simulates responses $(r_i)_{i=1}^t$ using comparisons $((\x_i, \x_i'))_{i=1}^t$.
For instance, if $\H$ is the class of linear functions, then $\text{fit}(\cdot, \cdot)$ can implement an SVM, logistic regression, or any other linear classification training procedure (with appropriate regularization).
Alternately, to rank cases 
based on the agent's responses (with $\H$ denoting the set of all rankings), the popular Bradley-Terry approach can be implemented within $\text{fit}(\cdot, \cdot)$
\cite{bradley198414}.
The choice of $\H$ here depends on prior beliefs about the agent's preference model.
However, a mismatch between the agent's preference model and $\H$ can impact the effectiveness of the framework (see Section~\ref{sec:model_misspecify}).

\section{Challenges to Modeling Moral Preferences} \label{sec:challenges}

{
Inspired by prior research from moral psychology, we highlight three obstacles to computationally modeling an agent's moral preferences.
These obstacles are (a) change in preference after making a certain number of decisions, (b) the agent's model not being included in $\H$, and (c) noise in the agent's responses.
We describe these challenges here, specifically focusing on prior empirical evidence for them from human subject research in the pairwise comparison setting.
}

{
\paragraph{Preference instability.} Empirical studies in psychology provide extensive evidence 
that agent's preferences in unfamiliar contexts are developed as they make decisions in those contexts \cite{hoeffler1999constructing,ariely2001timely,warren2011values,dhar1999comparison}.
In these settings, the first few choices made by an agent can be unstable (i.e., their preferences can change after making some decisions) and may not reflect their eventual preferences for future decisions.
Moral preferences can have similar instability and can be shaped by an agent's ongoing experience with the decision-making context
\cite{crockett2016computational,rehren2022stable, helzer2017once,curry2019mapping}.

In the context of pairwise comparisons, data from \citet{boerstler2024instability} provides evidence of this phenomenon in the kidney allocation setting.
In their study, participants are asked to participate in 10 sessions (one per day) and presented with 60 pairwise comparisons in each session. 
Session-specific analysis shows that, for many participants, there is significant variation in their weight distribution over the patient features across different sessions.
In other words, for many participants, their underlying utility functions change from session to session.
This kind of preference change can significantly impair the ability to computationally model moral preferences.

Note that we consider the instability of moral preferences over available actions and not preferences over moral values (e.g., one's value preference could be to prioritize equality in resource allocation over efficiency).
Moral values do inform moral judgments and the preferences an agent has over available actions.
But prior work has argued that while values are generally stable, agents can still be unstable in \emph{applying} those values to make moral judgments 
-- this is referred to as the ``value-action gap'' \cite{gould2023role}. 
In our setting, since we only observe the agent's moral judgments, we mainly focus on the challenge posed by the observed instability of preferences expressed through these judgments. 

\paragraph{Model misspecification.} Another challenge in the computational modeling of moral preferences is model misspecification, i.e., making incorrect/misrepresentative assumptions regarding the structure of the agent's decision-making process.
A popular modeling assumption is the \textit{additive independence} model, where we assume the utility the agent assigns to any input can be represented as a sum of the utilities assigned to individual input features \cite{chen2004survey} (e.g., linear utility satisfies this assumption).
Another common modeling assumption is the \textit{complete information assumption}, i.e., all the information \textit{explicitly used} by the agent to make their decision is available to the elicitation framework.
Assumptions of this kind are common in active learning-based elicitation as they reduce the complexity of query generation \cite{yang2021bayesian,johnston2023deploying}.
They also affect the choice of $\H$ in Algorithm~\ref{alg:pref_elicit}, e.g.,
if we assume additive independence and complete information, 
then setting $\H$ to be the linear class 
can help learn explainable representations of the agent's preferences.

However, in many situations, these assumptions do not reflect the agent's decision-making process \cite{pine2009encoding,gonzalez2021incomplete}.
Cognitive processes underlying moral decision-making are not clearly understood \cite{ugazio2022neuro} and can be more complex than a linear combination of available features \cite{hofmann2008comparison}.
Both empirical and theoretical analyses of moral judgments highlight this complexity.
\citet{cohen2016subjective} fit multiple kinds of linear and nonlinear models over people's responses in moral dilemmas and find that models from the exponential function family often provide the best fit.
\citet{kagan1988additive} theoretically questions both the additive and independence assumption (in an article appropriately titled ``The Additive Fallacy''), explaining through multiple \textit{contrastive} examples that (a) moral status of an act cannot always be determined by the sum of weights of individual features, and (b) weight assigned to each feature can depend on the weight assigned to other features (i.e., feature interactions).
As such, non-linearity and dependence across various features can be expected in moral decision-making processes.

\paragraph{Noisy responses.} Stochasticity in agent's choices,
(specifically, changes in their responses to similar scenarios at different times)
has been noted in various domains
\cite{marley2016choice,becker1963stochastic}.
The same is true for moral decision-making domains, where response variability can be a result of ongoing deliberation, increased decision ``difficulty'', and/or increased complexity of the decision context \cite{sivill2019ethical}.
\citet{boerstler2024instability} provide concrete evidence of this phenomenon.
In their kidney allocation study, participants take part in multiple sessions and six pairwise comparisons are repeated in each session.
Participants' responses to the repeated comparisons provide insight into response variability, quantified by the fraction of times a participant's choice to a repeated scenario differed from their \textit{majority choice} for this scenario.
\citet{boerstler2024instability} observe significant response variability for certain repeated comparisons (in the range of 10-18\%). 
Additionally, 
the results of \citet{boerstler2024instability} suggest that response variability is larger when the pairwise comparison is perceived as being more ``difficult'' by the participant, implying amplified stochasticity for difficult moral dilemmas.
}

\vspace{0.05in}
\noindent
All of these properties pose significant obstacles to the computational modeling of moral preferences. 
As we see in the following sections, the impact of these challenges can potentially be amplified by the use of active learning.

\section{Testing the Efficacy of Active Learning}

With the above challenges in mind, we next compare the performance of active learning for preference elicitation against the random baseline
over simulations of these challenges.

\paragraph{Simulation setup.} 
We primarily simulate agents that use linear utility functions, i.e.,
$u(\x) = \w^\top \x$, for any $\x \in \X$, given
weights $\w \in \R^d$.
The assumption of 
linear utility is quite prevalent in the preference elicitation literature (e.g., \citea{noothigattu2018voting}, \citea{McElfresh21:Indecision}, \citea{johnston2023deploying}).
In Section~\ref{sec:model_misspecify}, we will also question this assumption and simulate agents that use tree-based models and linear models with feature interactions.
To simulate an agent with linear utility, we sample weights $\w$ from the uniform distribution $\text{Unif}([-1, 1]^d)$.
We run Algorithm~\ref{alg:pref_elicit} for each simulated agent, presenting them with $N$ pairwise comparisons (ranging from 5 to 50). 
$\H$ is set to be the class of linear SVM classifiers over feature differences (with $\text{fit}(\cdot)$ performing SVM training).
Hence, each $h_t$ will contain the learned SVM weights, say $\hat{\w}_{h_t}$.
We evaluate performance using two metrics: (i) accuracy - for a held-out collection of 1000 comparisons, measure the fraction of comparisons for which the response using weights $\hat{\w}_{h_N}$ matches that of the agent -  and (ii) normalized distance - measure the $L_2$-distance between $\hat{\w}_{h_N}$ and $\w$ after normalization.
For each setup, we report the mean and standard deviation of these metrics across 50 simulated agents.
In the main body of the paper, we will primarily discuss the accuracy metric. Results with respect to distance are similar but deferred to Appendix~\ref{sec:add_results}.
The number of features $d$ is varied from $\set{3, \dots, 15}$ and each feature has range $\set{1, \dots, 10}$ (unless specified otherwise).
\textsc{Active-VS-PE} and \textsc{Active-Bayes-PE} will use a linear kernel function $\kappa$.
Other implementation details are presented in Appendix~\ref{sec:impl_details}.

\begin{figure*}[t]
    \centering
    \includegraphics[width=\linewidth]{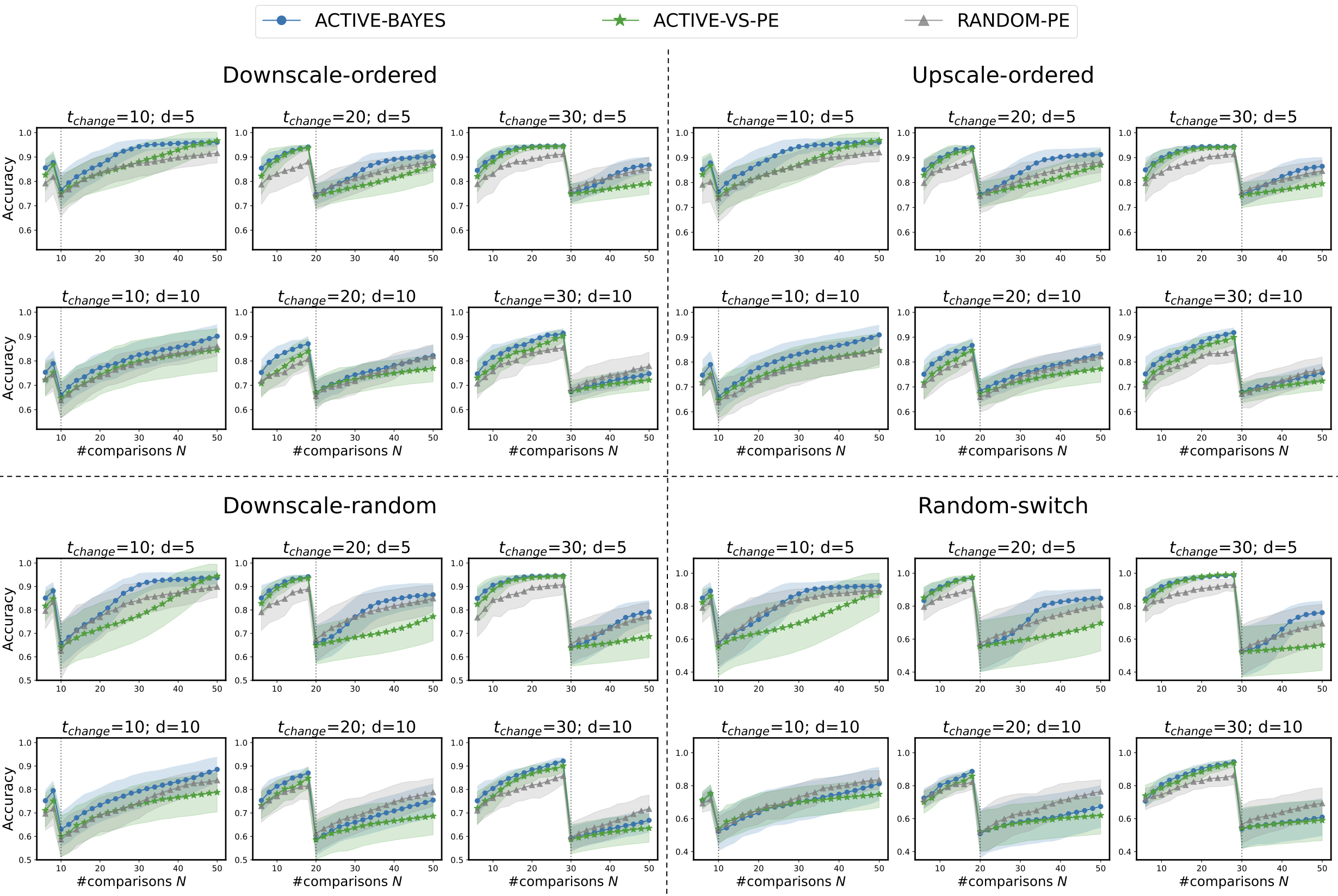}
    \caption{Performance for preference-change scenarios from Section~\ref{sec:pref_change}. \textsc{Active-Bayes-PE} often performs better than \textsc{Random-PE} post-$t_{\text{change}}$ when $d{=}5$. However, in many cases (e.g., $d{=}10$, $t_{\text{change}}{=}20, 30$), both active learning algorithms have similar or worse performance than \textsc{Random-PE}.}
    \label{fig:pref_change_combined}
\end{figure*}

\subsection{Preference Instability} \label{sec:pref_change}
The first challenge we discuss in Section~\ref{sec:challenges} is preference instability, i.e., the agent's underlying preferences can change after making some decisions.
Since the next query suggested by active learning depends on the agent's responses to comparisons presented so far, we simulate scenarios where an agent's preferences undergo changes to assess the impact of preference instability on active learning algorithms.

    We assume that the agent's utility function is linear.
    Suppose that the agent changes their preferences once, at timestep  
    $t_{\text{change}} \in [N]$.
    Let $w^{\text{pre}} \in R^d$ denote the agent's weight vector for all timesteps $t < t_{\text{change}}$ and $w^{\text{post}} \in R^d$ denote the agent's weight vector for all timesteps $t \geq t_{\text{change}}$.
    We simulate the following kinds of preference changes.
    % 
    % 
    % \item 
    \begin{itemize}
    \item \textbf{Downscale-ordered.} Agent changes their preference utility function to only use the feature to which they assigned the highest weight previously.
    For this agent, we sample pre-change preference $w^{\text{pre}} \sim \text{Unif}([-1, 1]^d)$ and set $I=\arg\max_{i} |w^{\text{pre}}_i|$. Then, for post-change preference, $w^{\text{post}}_I = w^{\text{pre}}_I$ and $w^{\text{post}}_i = 0$ for all $i \in [d]\setminus\set{I}$.    
    \item \textbf{Downscale-random.} Agent changes their utility function to again use only one feature, but the feature is randomly selected.
    For this agent, we sample pre-change preference $w^{\text{pre}} \sim \text{Unif}([-1, 1]^d)$ and set $I$ is chosen randomly from set $[d]$. Then, for post-change preference, $w^{\text{post}}_I = w^{\text{pre}}_I$ and $w^{\text{post}}_i = 0$ for all $i \in [d]\setminus\set{I}$.    
    % 
    
    % \item 
    \item \textbf{Upscale-ordered}: Agent changes preference utility function from using just one feature to all features,
    with features in $w^{\text{post}}$ having lower relative weight than the non-zero weight in $w^{\text{pre}}$.
    For this agent, sample $w^{\text{post}} \sim \text{Unif}([-1, 1]^d)$ and $I = \arg\max_{i} |w^{\text{post}}_i|$. Then, $w^{\text{pre}}_I =w^{\text{post}}_I$ and $w^{\text{pre}}_i=0$ for all $i{\in}[d]\setminus\set{I}$.   
    % 
    % \item 
    
    \item \textbf{Random-switch.} Agent changes to a random new preference after  $t_{\text{change}}$.
    Here, we sample both weights vectors $w^{\text{pre}}, w^{\text{post}} \sim \text{Unif}[-1, 1]^d$, {independently.}
    \end{itemize}

\noindent
\textbf{Downscale-ordered} and \textbf{Downscale-random} model the settings where an agent changes their preference to reduce decision-making effort
\cite{shah2008heuristics}.
In certain cases, the agent can choose only to use the feature that was most important to them pre-$t_{\text{change}}$, which is modeled by \textbf{Downscale-ordered}.
\textbf{Upscale-ordered} is a symmetric scenario where the agent instead
incorporates additional features in their preference.
Finally, \textbf{Random-switch} models agents who make more drastic changes to their preference, e.g. following an entirely different set of moral norms.
Appendix~\ref{sec:add_results_pi} models multiple other scenarios as well, e.g., agent downscaling/upscaling to or from random features (instead of highest weighted feature) and downscaling/upscaling to or from more than one feature.
For all scenarios, we compare the accuracy achieved by \textsc{Active-VS-PE} and \textsc{Active-Bayes-PE} vs.\   \textsc{Random-PE}, varying the number of features, number of comparisons, and $t_{\text{change}}$.

\paragraph{Results.}
The results of our simulations are presented in Figure~\ref{fig:pref_change_combined}.
As expected,  \textsc{Active-VS-PE} and \textsc{Active-Bayes-PE} always achieve higher accuracy than \textsc{Random-PE} prior to $t_{\text{change}}$.
Post-$t_{\text{change}}$ performance shows how well each algorithm recovers from preference change.

Let us first look at the \textbf{Downscale-ordered} setting
(plots on the top-left side of Figure~\ref{fig:pref_change_combined}).
In this case, when the preference change occurs early (i.e., $t_{\text{change}} = 10$), 
\textsc{Active-Bayes-PE} recovers quite fast from the preference change: the accuracy of \textsc{Active-Bayes-PE} becomes higher than that of \textsc{Random-PE} within 10 timesteps (on average) post-$t_{\text{change}}$ when $d=5$.
In comparison, \textsc{Active-VS-PE} takes longer to recover and exceeds \textsc{Random-PE} in accuracy.
For larger $t_{\text{change}}$, both active learning approaches seem to
recover slower and incompletely. 
When $d=10$ and $t_{\text{change}}$ is 20 or 30, we further observe that \textsc{Active-Bayes-PE} and \textsc{Active-VS-PE} have similar or even lower accuracy than \textsc{Random-PE} for all timesteps post-$t_{\text{change}}$. 
This also implies reduced efficiency of active learning in settings with high feature complexity; 
for any desired level of accuracy, active learning approaches take a similar or larger number of comparisons than the random query baseline to achieve that accuracy level.
In the case of \textbf{Downscale-random} setting, the performance of active learning algorithms, relative to \textsc{Random-PE}, follows similar patterns -- when $d=5$ and $t_{\text{change}}$ is small, \textsc{Active-Bayes-PE} recovers well compared to other algorithms, but this recovery is much slower for $d=10$.
The drop in accuracy around timestep $t_{\text{change}}$ is also larger in magnitude for \textbf{Downscale-random} compared to \textbf{Downscale-ordered}; this is expected since there is relatively more consistency between pre-change and post-change preferences in the \textbf{Downscale-ordered} setting.

Similar trends are observed for the
\textbf{Upscale-Ordered} and \textbf{Random-switch} plots in Figure~\ref{fig:pref_change_combined}.
Active learning approaches have the worst recovery 
in the \textbf{Random-switch} setting where, due to the drastic change in the agent's preferences, both \textsc{Active-BAYES-PE} and \textsc{Active-VS-PE} have similar or worse performance than the \textsc{Random-PE} post $t_{\text{change}}$ when $d=10$.
On the positive side, when $d=5$, \textsc{Active-BAYES-PE} does 
achieve higher accuracy than \textsc{Random-PE} within 20 timesteps post-$t_{\text{change}}$ on average.

Overall, Bayesian active learning approaches can efficiently elicit preferences while handling preference changes when  
the number of features $d$ is small.
However, these approaches fail to provide similarly improved performance as feature complexity and preference change timestep increases.
These results highlight the importance of knowing the nature and scale of preference instability before deploying active learning.
While active learning will eventually recover
after a larger number of timesteps beyond 50, we see that in the timesteps following $t_{\text{change}}$, it can perform even worse than the random baseline due to its dependence on the agent's previous responses.
Considering that active learning is usually employed when one has to be economical with the number of presented comparisons (due to time and/or cost constraints), not being able to rely on a certain number of initial responses can significantly affect the accuracy of the learned preferences and fail to improve, or even harm, the efficiency of the framework.

\subsection{Model misspecification} \label{sec:model_misspecify}

The second challenge we discussed in Section~\ref{sec:challenges} is model misspecification, specifically questioning the \textit{additive independence} and \textit{complete information} assumptions for the agent's moral decision-making process.
In this section, we evaluate active learning when additive independence and complete information assumptions are not satisfied.
Setting $\H$ to be the linear class,
we simulate the following scenarios.

\begin{itemize}
    \item 
    \textbf{Agent uses tree-based utility.} 
    We simulate agents that use shallow binary decision trees to assign utility.
    Tree-based models reflect decisions made
    using \textit{if-then} rules; e.g.,
    in an organ allocation setting, an agent might assign a higher utility to a patient if their age is ${>}50$ but can be indifferent to the exact age number.
    To maintain parity between \textit{capacity} of a tree model and models in $\H$,
    we simulate agents with tree models of depth $\lfloor \log d \rfloor$, where $d$ is the number of features.
    We simulate this scenario with binary and non-binary features.

    \item
    \textbf{Agent uses second-order interaction terms.} 
    Even with a linear utility model, the agent's utility function could use interactions between different features.
    Interaction terms
    account for scenarios where the importance an agent might assign to any feature is correlated with the value of another feature.
    For example, in the organ allocation setting, an agent might assign a higher weight to a patient's number of dependents if the patient is young, implying an interaction between the age and number of dependents variables.
    We simulate this scenario by measuring performance across a varying number of features $d$ and a varying number of second-order interactions.

    \item \textbf{Missing features.} Finally, we consider the scenario where the agent uses information unavailable to the elicitation framework.
    We simulate this scenario by allowing the agent to use a larger feature set than that available for elicitation.
    Our simulations assess performance across a varying number of total and missing features.
\end{itemize}

\begin{figure}[t]
    \centering
    \includegraphics[width=0.8\linewidth]{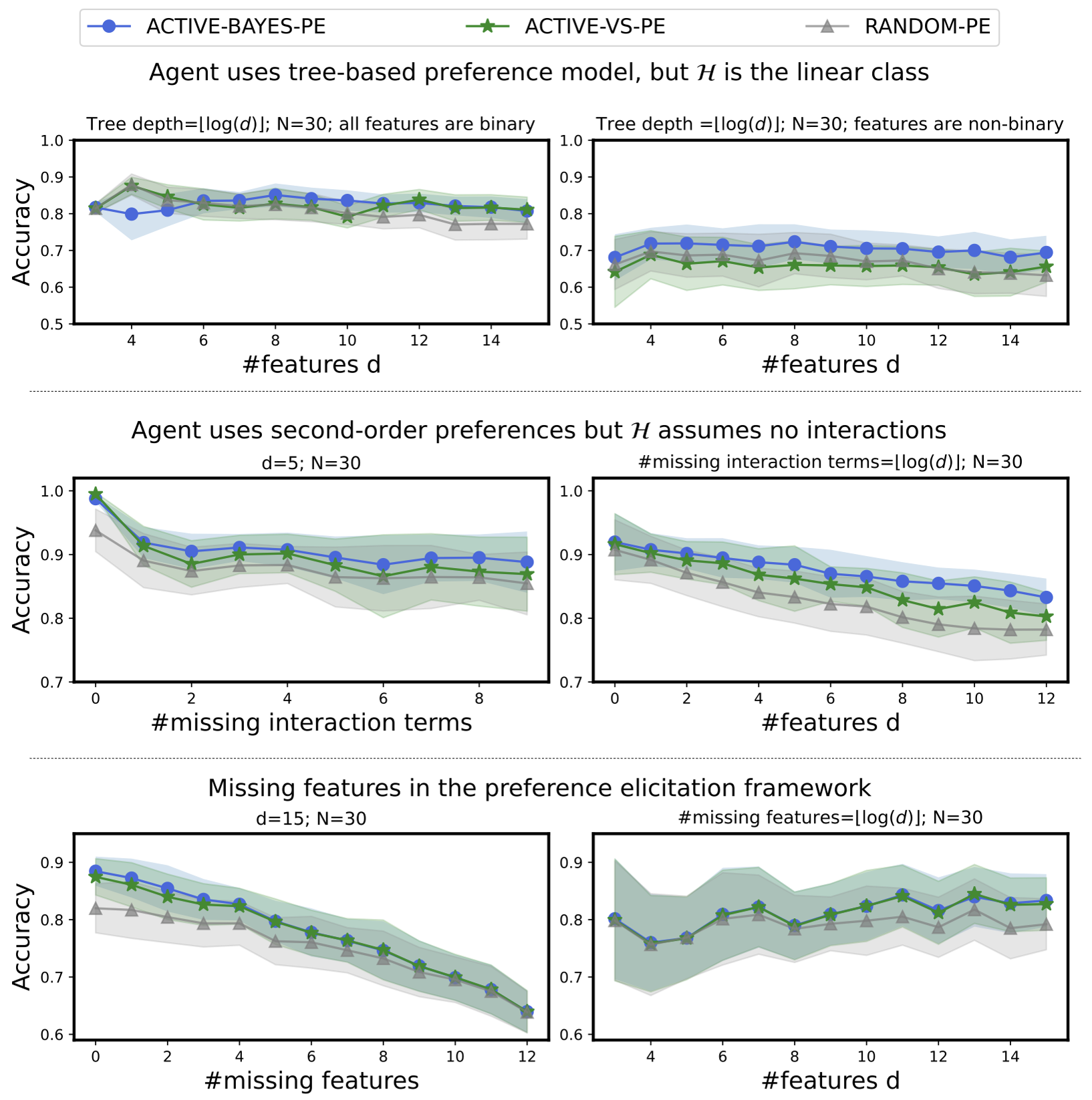}
    \caption{Performance for model misspecfication scenarios from Section~\ref{sec:model_misspecify}.
    Active learning is more effective when the extent of model misspecification is small in scale.
    }
    \label{fig:model_misspecify_comb}
\end{figure}

\begin{figure*}
    \centering
    \includegraphics[width=\linewidth]{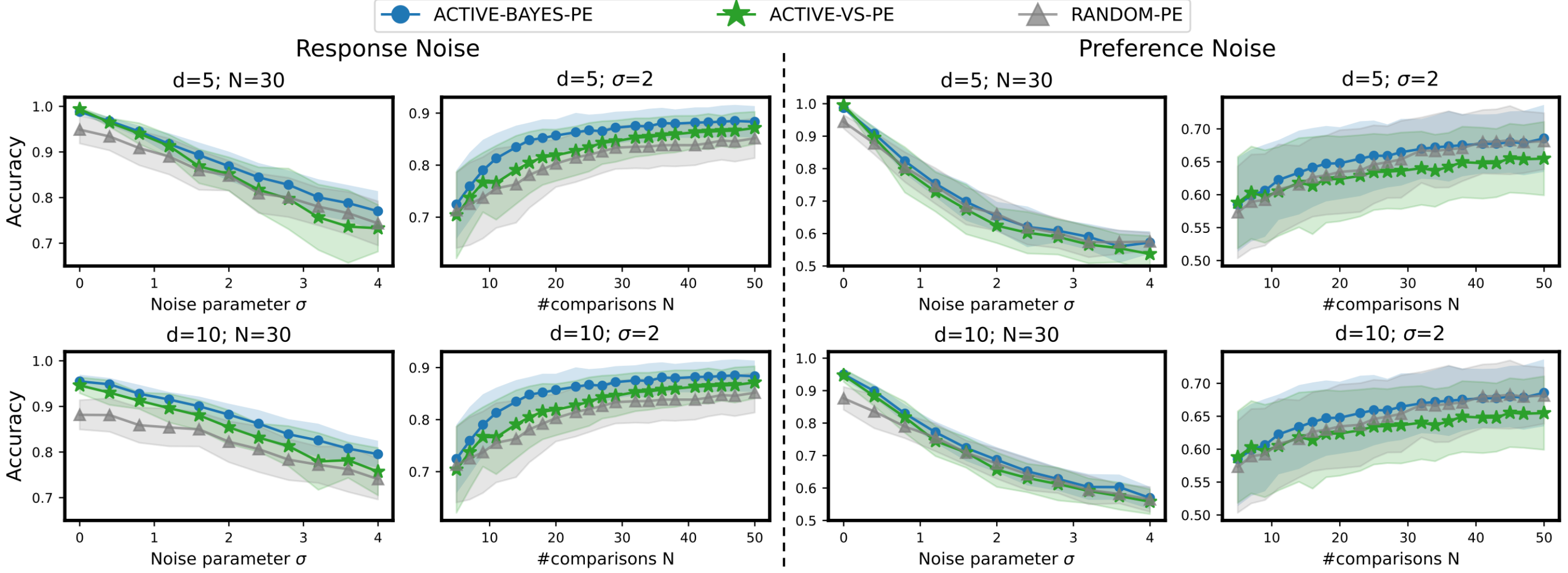}
    \caption{Performance for the noise models from Section~\ref{sec:noise_analysis}. \textsc{Active-BAYES-PE} performs better than the random query baseline even with response noise. However, it fails to provide a similar improvement 
    in most scenarios of preference noise.}
    \label{fig:noise_results_combined}
\end{figure*}

\noindent
\textbf{Results.}
The results for these simulations are presented in Figure~\ref{fig:model_misspecify_comb}.
When the agent uses tree-based preference, \textsc{Active-Bayes-PE} has marginally better accuracy than the \textsc{Random-PE} after 30 comparisons when $d$ is large.
For small $d$, both active learning approaches tend to have similar or worse accuracy than the random baseline.
The impact of model misspecification 
also depends on the input domain -- overall accuracy is lower for non-binary features.

When the agent uses interaction terms, Figure~\ref{fig:model_misspecify_comb} shows that accuracy decreases as the number of interaction terms increases.
% leads to a decrease in accuracy.
However, when the number of interaction terms is much smaller 
than $d$, \textsc{Active-Bayes-PE} and \textsc{Active-VS-PE} can achieve higher accuracy than \textsc{Random-PE} after 30 comparisons.
Finally, in the case of missing features, the larger the number of missing features (relative to $d$), the lower the accuracy, and the smaller the gap between \textsc{Active-Bayes-PE}, \textsc{Active-VS-PE}, and \textsc{Random-PE} after 30 comparisons. 
Missing information reduces the capacity of the framework to capture the agent's decision-making process, leading to an accuracy drop.

For these scenarios, we see that the larger the scale of disparity between the agent's utility function and $\H$,
the worse the performance of active learning as compared to the random query baseline.
Active learning might still converge to the best hypothesis in $\H$ (see accuracy vs.\ timestep results in Appendix~\ref{sec:add_results_mm}); however, the above results show that disparity between functions in $\H$ and the agent's utility affects active learning's ability in generating informative queries and leads to a reduction in accuracy of learned preferences.

\subsection{Noisy Responses} \label{sec:noise_analysis}
The final challenge we highlighted in Section~\ref{sec:challenges} is stochasticity or variability in agent's responses to moral dilemmas.
Two ways in which this stochasticity has been modeled in prior literature are (a) \emph{response noise}: noise that arises and affects the agent's response after the agent has computed utility for the presented cases,
and (b) \emph{preference noise}: noise that arises due to variability in the agent's underlying utility function \cite{bhatia2017noisy, marley2016choice}.
Suppose the agent uses linear utility, i.e., 
$u(\x) = \w^\top \x$, for some $\w{\in}\R^d$.
Then, the above noise models can simulated as follows.

\begin{itemize}
    \item 
    \textbf{Response noise model.} This model induces noise $\varepsilon \sim \mathcal{N}(0, \sigma^2)$ after utility is computed. 
    Assuming an additive noise model, the impact of this noise on 
    the agent's response $R$ can be interpreted as
    changing it to
    $R(\x, \x') = \mathbf{1}[u(\x) - u(\x') + \varepsilon > 0].$
    Our simulations evaluate performance for varying $\sigma \in \R$.
    
    \item  \textbf{Preference noise model.} This model assumes noise in the utility generation process itself. 
    We simulate this setting as follows:
    Suppose that whenever presented with a pairwise comparison,
    the agent first samples $\w \sim \mathcal{N}(\w^\star, \sigma^2\mathbf{I}/d)$, and then uses the sampled $\w$ to compute utilities.
    Here, $\w^\star \in \R^d$ represents summary feature weights assigned by the agent and $\sigma \in \R$ is the noise parameter varied in our simulations.
 \end{itemize}

\noindent
\textbf{Results.} The results for this simulation are presented in Figure~\ref{fig:noise_results_combined}.
As expected, increasing $\sigma$ leads to a decrease in accuracy of all algorithms.
However, in the case of response noise, \textsc{Active-Bayes-PE} has higher accuracy than the \textsc{Random-PE} baseline even for high values of $\sigma$.
Accuracy vs number of comparisons plot for $\sigma = 2$ further shows that \textsc{Active-Bayes-PE} starts achieving higher accuracy than \textsc{Random-PE} with as few as 20 comparisons.
Performance of \textsc{Active-VS-PE}, on the other hand, is relatively better than \textsc{Random-PE} for small $\sigma$ values but becomes similar to that of \textsc{Random-PE} for large $\sigma$.
Hence, in this case, active learning (especially, \textsc{Active-Bayes-PE}) can be relatively more accurate at preference elicitation despite noise.

In the preference noise setting, both active learning approaches have similar performance as the \textsc{Random} baseline for almost all non-zero $\sigma$ values.
Variation with respect to $\sigma$ and number of comparisons shows that noise in preference weights significantly affects the ability of all algorithms to learn the underlying preferences when $\sigma{>}1$.
Hence, here active learning fails to provide any performance improvement in comparison to the random query generation baseline.

\section{Discussion, Limitations, and Future Work}

Through the presented simulations, we 
highlight how potential issues associated with moral preferences, such as preference instability, response variability,
or modeling errors, can 
impact the efficacy of active-learning-based preference elicitation.
In all simulated scenarios, we compare the performance of active learning-based preference elicitation against the baseline method of using random queries at each time step.
Overall, there are \textit{positive scenarios} where active learning still performs better than the random query baseline -- e.g., when noise affects utility but not the underlying preferences, or in the case of small-scale preference instability in initial iterations.
Then, there are \textit{neutral scenarios} where the simulated challenge impacts the efficiency of all algorithms similarly and the performance of active learning and the random baseline are comparable -- e.g., for large-scale modeling errors or when the agent's underlying preferences are noisy.
In these cases, using active learning does not provide any added benefit but it also does not cause any harm to the elicitation framework.
Finally, there are \textit{negative scenarios}, where using active learning is less effective than the random baseline 
-- e.g., when the number of features is large and the agent's preference changes after they have responded to a large number of comparisons.
Here, since active learning uses the agent's previous responses to construct the next query, it takes longer to recover from preference changes.

Different real-world challenges have different effects on active learning for preference elicitation.
Deploying these frameworks without prior understanding of the agent's decision-making for the given context can lead to inaccurate representations of their preferences.
While using a small number of queries will almost always provide only an approximate representation of the underlying preferences, our simulations call attention to the sources of inaccuracy 
that were unappreciated in previous works and could lead to incorrect interpretations of results if not considered in practice.

In the paragraphs below, we highlight other characteristics of our assessment as well as future work on this topic.

\paragraph{Algorithmic solutions.}
One response to the challenges we simulate is that many of them can be addressed algorithmically if they are known in advance.
If an agent's preferences are known to be unstable for initial comparisons, then one can,
say, modify the elicitation approach to disregard a certain number of initial comparisons
or assign sample weights to each case that are inversely proportional to the duration since the case was observed by the agent.
This way, active learning can construct queries that are primarily based on the most recent agent responses.
To account for feature interactions, the models in $\H$ can allow interactions by default and use regularization
to rule out scenarios where interactions are not used.
Prior work on active learning methods that are robust to noise or distribution shifts can be potentially adapted to make elicitation more resilient to noise or modeling errors \cite{angluin1988learning, zhao2021active}.
In simulations, Bayesian approaches often appear more robust to certain challenges, e.g., small-scale instability.
Hence, one approach is to use \textsc{Active-bayes-PE} with an expanded hypothesis class $\H$ (e.g., combining linear and tree classes) to counter issues of model misspecification. The main challenge here is creating an efficient query-selection algorithm over an expanded $\H$ while being robust to instability and noise, and can be explored as part of future work.
Most of these modifications, however, require prior knowledge of the nature of the challenge associated with the agent's decision-making process.
Indeed, the primary goal of our analysis is to highlight that certain assumptions made when using active learning incorrectly rule out these challenges.
Knowing that these assumptions might be violated can help practitioners develop modifications that might be better suited for the given context.
Also, some active learning algorithms may be generally more robust to violated assumptions than others.

\paragraph{Sensitivity of moral preferences.}
As discussed, moral preferences can be different from generic preferences for self-benefit and, hence, assuming moral preferences to have a similar structure as other preferences will hurt the accuracy of the elicitation framework.
Based on prior insights from the literature on moral preferences, our work discusses specific mechanisms via which these inaccuracies can occur.
With the highlighted challenges and considering the emergent nature of moral psychology research, % or new/in early stages
the task of eliciting moral preferences can be tricky.
Nevertheless, building elicitation methods specifically for moral preferences is a worthwhile direction for future research, given their role in creating ethical AI tools.
\vk{At the same time, 
moral preference elicitation is just one (albeit complex) part of ethical AI development.
Mechanisms to incorporate learned moral preferences within AI systems 
involve additional work and should be similarly subjected to technical analyses of feasibility under various real-world challenges.}

\paragraph{On utility functions.}
\vk{Our framework employs utility functions to model people's preferences over actions in moral dilemmas, as is standard practice in this literature. Despite the overlap in naming conventions, it is important to clarify that modeling moral preferences using utility functions does not presuppose a reliance on utilitarian or consequentialist moral theories (as long as consequentialism isn't used generically to cover all possible theories \cite{portmore2022consequentializing}). The justifications people have for considering features that contribute to their utility function do not have to draw on consequentialist principles, and the features people consider may not impact future consequences directly, such as when people think patients' past criminal behavior is important for determining who should receive an available kidney.  
Adherence to many different moral theories (including non-consequentialist theories) can be modeled using utility functions, and our analysis aims to call attention to challenges that can arise when using active learning to obtain accurate representations of various utility functions. 
Nevertheless, future work is needed to assess the effectiveness and challenges of using active learning to predict moral judgments under other modeling frameworks or conditions, e.g., when using explicit moral constraints \cite{black2020absolute}, harm-based utilities \cite{beckers2022causal}, or modified utility-based frameworks that explicitly account for deontological values \cite{lazar2017deontological}.
}

\paragraph{On non-moral preferences.} 
{
Issues of instability, noise, or model misspecification can arise with non-moral preferences as well.
Yet, we focus on moral preferences because the specific challenges we simulate are inspired by the literature on moral philosophy and psychology. AI applications that would rely on moral preference elicitation often involve high stakes and errors in preference elicitation can cause undue harm to users and impacted individuals (e.g., in autonomous vehicles and kidney allocation settings), requiring high levels of elicitation accuracy and reliability. 
}

\paragraph{Other analyses/baselines.} 
{
Future assessments of active learning can also simulate violations of multiple assumptions; e.g., the presence of both preference instability and model errors. These combinations can be reflective of more complex decision-making settings.
Additionally, in applications where data from past agents is available,
other baselines
(beyond simple random query baseline) 
can be considered.
For instance, one could create 
elicitation using a curated set of queries that were informative of the preferences of past agents.
All or random subsets of this curated set can be used to elicit preferences.
Evaluation of active learning against such baselines can provide insight into whether it is better than methods that use prior information.
}

\paragraph{Limitations of our analysis.}
Our simulations demonstrate the need for improved modeling of human moral preferences and developing active learning approaches that are more robust to the simulated challenges.
Along with this direction for future work, additional analyses can be conducted to further discover other failure points of quantitative preference elicitation frameworks.
Note that all of our analysis simulates agents with linear or tree-based utility functions. 
Human moral preferences can be more complex and analyzing active learning performance through real-world data can provide more robust results.
In particular, this will require human-subject studies where participants respond to comparisons generated using active learning and random comparisons.
% (similar to the ones performed in \citet{boerstler2024instability}).
As expected, collecting this data will be expensive and time-consuming.
In that regard, 
our simulation provides a starting point on the kind of data that can be gathered using active learning and raises challenges that need to be accounted for when analyzing this data.

\section{Conclusion}
The results of our simulations highlight the challenges associated with extracting accurate representations of agents' moral preferences while using as few queries as possible.
In cases of large-scale instability or noise in agent preferences or responses, active learning has similar or worse performance than the random baseline.
The assumptions made by the elicitation framework regarding the agent's moral preferences also impact the effectiveness of active learning.
The use of active learning for moral preference elicitation therefore requires careful evaluation of modelling assumptions and the scale of expected variability in agent preferences and responses for the relevant context.
If large-scale instability, noise, and/or violation of modeling assumptions are expected, then
appropriate alternatives or modifications to active learning should be considered to counter such issues.

\section*{Acknowledgements}

We are grateful for financial support from OpenAI \& Duke.

\clearpage
\bibliography{references.bib}

\clearpage

\appendix

\section{Details of Active Learning Algorithms} \label{sec:active_descriptions}
In this section, we provide additional details of the two active learning algorithms that we evaluate in the main body.

\paragraph{Version Space-based Active Learning -- \textsc{Active-VS-PE}.} 
The first approach relies on kernel SVM-based classification. 
Given an SVM decision boundary learned using available labeled comparisons for an agent, the informativeness of any new query can be quantified using the distance of the query from the decision boundary and this heuristic can be used to generate an informative next query for the agent.
For any given input $\x \in \R^d$, an SVM classifier computes
\[f_\w(\x) = \ip{\w, \phi(\x)},\]
where $\phi : \R^d \rightarrow \F$ is a mapping to a kernel-induced space $\F$ for a given kernel $\kappa: \R^d \times \R^d \rightarrow \R$, such that $\kappa(\x, \x') = \ip{\phi(\x), \phi(\x')}$, 
$\w \in \F$ is the weight vector, and the classifier decision is 1 if $f_\w(\x) > 0$ and $-1$ otherwise\footnote{Since we are dealing with pairwise comparisons, we will not include any bias term in the SVM functional form.}.
Suppose we have a labelled training set $S = \set{(\x_i, y_i)}_i$, where $\x_i \in \R^d$ and $y_i \in \set{1, -1}$.
The \textit{version space} for $S$ denotes all vectors $w \in \F$ that fit $S$, i.e., 
\[\V(S) := \set{\w \mid \norm{\w}=1, y\cdot f_\w(\x) > 0 \text{ for all } (\x,y) \in S}.\]
The goal of active learning here is to select an element to add to $S$ which significantly reduces the size of $\V(S)$.
To that end, \citea{tong2001support} propose sampling a query that is closest to the decision boundary, i.e,
sampling an element $\hat{\x} :=  \arg\min_u |\ip{\w, \phi(\x)}| = \arg\min_u |f_\w(\x)|$.\footnote{Alternative SVM-based sampling heuristics that can have better theoretical and empirical real-world performance have also been proposed in other works \cite{kremer2014active}. We tested these heuristics in our simulation setting and they have similar performance as the one described above.}
To implement this approach in Algorithm~\ref{alg:pref_elicit}, at each timestep $t$, we are given the
dataset $((\x_i, \x_i'))_{i=1}^t$ to labels $(r_i)_{i=1}^t$.
Hence, sampling using the above SVM-based approach would require first learning an SVM hypothesis $f$ that best fits  $((\x_i, \x_i'))_{i=1}^t$ to labels $(r_i)_{i=1}^t$ and then finding the pairwise comparison $\arg\min_{(\x, \x')} |f((\x, \x'))|$.

\paragraph{Bayesian Active Learning -- \textsc{Active-Bayes-PE}.} 
Another popular active learning-based preference elicitation approach uses a Bayesian framework for sampling based on information learned from previous agent responses.
% % 
The Bayesian Active Learning with Disagreement (BALD) algorithm by \cite{houlsby2011bayesian} models individual preferences using a Gaussian process with a specified kernel.
Based on the Gaussian Process classification literature, suppose each point $\x \in \R^d$ can be characterized by a latent value,  $f(\x)$, such that $f$ follows a Gaussian distribution. That is,
$f \sim GP(\mu(\cdot), \kappa(\cdot, \cdot)),$ where $\mu$ denotes the mean function and $\kappa$ is a pre-defined kernel.
We can model the label $y \in \set{0,1}$ at any point $\x$ as following the Bernoulli$(\Phi(f(\x)))$ distribution, where $\Phi$ is the Gaussian CDF function.
In this setup, suppose the following queries have been made so far: $S : \{(\x_i, y_i)\}_{i=1}^t$.
Then, the BALD approach suggests selecting the next query to be the one that maximizes mutual information or the decrease in expected posterior entropy, i.e.,
\[\arg\max_{\x} H(y \mid \x, S) - \E_{f \sim P[f \mid S]}[H(y \mid \x, f)].\]
\citea{houlsby2011bayesian} show that, in case of Gaussian process prior for the function $f$, we can approximate the quantity $H(y \mid \x, S) - \E_{f \sim P[f \mid S]}[H(y \mid \x, f)]$, for any query $\x$, as 
\begin{equation}
\label{eq:entropy_factor}    
I(\x) := h\left( \Phi \left( \frac{\mu_{\x, S}}{\sqrt{\sigma^2_{\x,S} + 1}} \right) \right) - C \frac{\exp\left(-\frac{\mu^2_{\x, S}}{2(\sigma^2_{\x,S} + C^2)}\right)}{\sqrt{\sigma^2_{\x,S} + C^2}}.
\end{equation}
Here $h(p) := -p\log p - (1-p)\log(1-p), C = \sqrt{\pi \ln 2/2}$,  and $\mu_{\x, S}, \sigma_{\x, S}$ are the posterior predictive mean and deviation of $f(\x)$.
Hence, in the active learning setup, the goal in every iteration is to find the query $x$ that maximizes $I(\x)$.

Implementation of this approach for the pairwise comparison setting of Algorithm~\ref{alg:pref_elicit} first requires learning a representation of the posterior of $f(\cdot)$ corresponding to the labeled dataset $((\x_i, \x_i'))_{i=1}^t$, $(r_i)_{i=1}^t$ and then computing a new pairwise comparison $(\x, \x')$ that maximizes $I((\x,\x'))$.

\paragraph{Use-cases of the above algorithms.} 
% \todo{...}
The two active learning approaches we consider represent two different paradigms for query sampling.
\textsc{Active-VS-PE} is a non-probabilistic discriminative sampling method while \textsc{Active-Bayes-PE} represents a probabilistic entropy-based approach.
Depending on the application, one might be favored over the other.
The Bayesian can be favorable when the information gain function $I(\cdot)$ is submodular -- allowing for near-optimal optimization using greedy approaches.
Another advantage of this approach is that it is independent of model choices or optimization methods. 
The version-space approach, on the other hand, is easier to implement and performs better for certain applications like text classification.
However, there are theoretical similarities between the two approaches and they can be equivalent in some settings as well \cite{houlsby2011bayesian}.

\begin{figure*}[!hbtp]
    \centering
    \includegraphics[width=\linewidth]{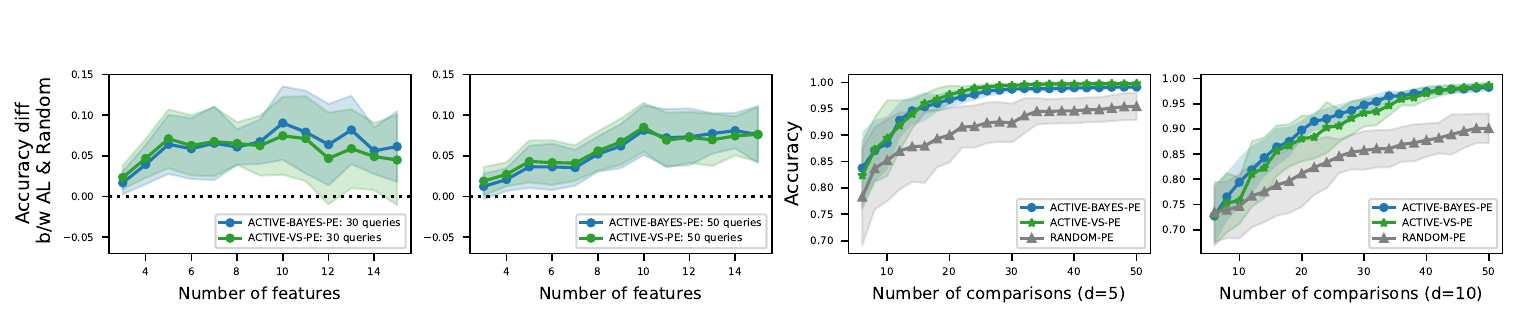}
    \caption{Performance of \textsc{Active-VS-PE}, \textsc{Active-Bayes-PE} and \textsc{Random-PE} in an ``idealized setting'' (i.e., no assumption violations).}
    \label{fig:active_ideal_performance}
\end{figure*}

\section{Additional Implementation Details} \label{sec:impl_details}

Implementation details of the presented simulations that were excluded from the main body are presented here.

\paragraph{Other details of \textsc{Active-Bayes-PE} and \textsc{Active-VS-PE} implementation.}
As mentioned earlier, both \textsc{Active-Bayes-PE} and \textsc{Active-VS-PE} use linear kernels.
For each method, at every time-step $t$, we sample 1000 new comparisons $T := \set{(\hat{\x_i}, \hat{\x_i}')}_i$ and compute the \textit{informativeness} of each comparison as defined by the method.

For \textsc{Active-VS-PE}, we first learn an SVM hypothesis (using the labelled comparisons), say with coefficients $\hat{\w}_t$,
and
measure the absolute value of dot product between feature differences for each comparison in $T$ and $\hat{\w}_t$, i.e., we compute $\set{|(\hat{\x_i} - \hat{\x_i}')^\top \hat{\w}_t|}_i$. The chosen comparison is the one with the smallest absolute dot product value.

For \textsc{Active-Bayes-PE}, when using a linear kernel, we first employ Bayesian ARD regression (over the labelled comparisons) with priors for the regularization and precision parameters set to $\Gamma(1,1)$.
With the kernel parameters obtained from this regression, we compute the mean and variance of the latent value for each pairwise comparison difference in $T$.
For each $(\hat{\x}, \hat{\x}') \in T$, we then compute $I(\hat{\x} - \hat{\x}')$ (where $I(\cdot)$ is the mutual information function and defined in the description of the bayesian active learning approach), and choose the comparison that maximizes this value.
% The chosen comparison is the one with the maximum value of $I(\hat{\x_i} - \hat{\x_i}')$.
% 
% 
We also tested \textsc{Active-Bayes-PE} with the standard RBF kernel and it had a similar or worse performance than the linear kernel approach.

\paragraph{Model fitting details.}
The hypothesis class $\H$ is set to be the class of linear SVM functions and the fit$(\cdot, \cdot)$ function executes the standard linear-SVM training procedure (using python sklearn library) with $L_2$-regularization.
Specifically, we set these functions to operate over feature differences, i.e., for any given comparison $(\x, \x')$, the input to an SVM classifier $h_t$ in the feature-wise difference $(\x - \x')$.
Using feature differences is a standard approach in pairwise comparison setting in practice \cite{freedman2020adapting}.
Correspondingly, the vector $\hat{w}_{h_t}$ for each $h_t$ represents weights assigned to individual feature differences.
% 
% We use the python sklearn library to

\paragraph{Implementing the tree-based utility model.}
Finally, to simulate an agent with a tree-based model of depth $d'$, we essentially create a binary tree of depth $d'$, choosing random features and feature values for partitioning at each node.

\section{Additional results} \label{sec:add_results}

In this section, we provide additional results that had to be omitted from the main body due to space constraints.

First, Figure~\ref{fig:active_ideal_performance} denotes the performance of both active learning algorithms in an ``ideal'' setting, i.e., when preferences are stable and do not suffer from any kind of noise and when both the underlying preferences and the hypothesis class $\H$ are linear functions over the available features (without interactions).
In this case, we clearly see the advantage of using active learning. Both \textsc{Active-Bayes-PE} and \textsc{Active-VS-PE} outperform the \textsc{Random-PE} baseline in all cases.

\subsection{Preference Instability} \label{sec:add_results_pi}

\paragraph{Performance with respect to normalized distance metric.}

\begin{figure*}[t]
    \centering
    \includegraphics[width=\linewidth]{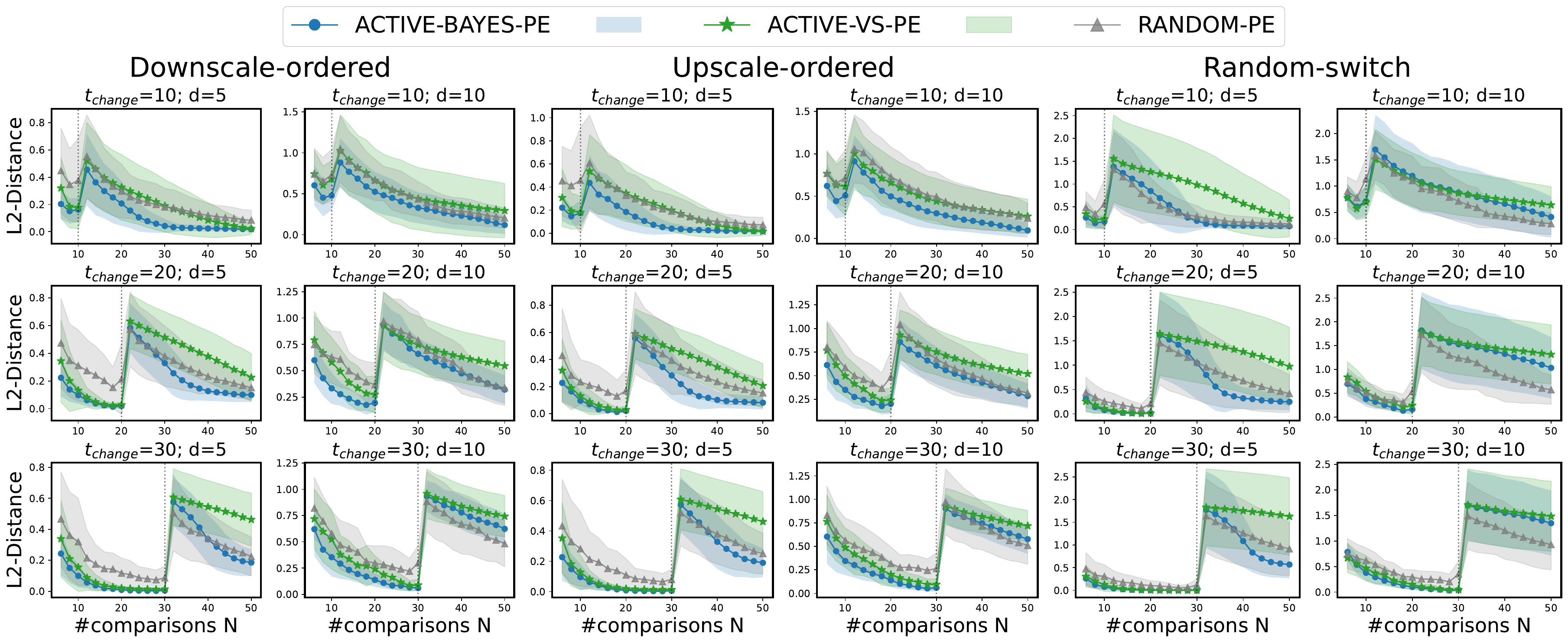}
    \caption{Normalized L2-distance vs number of comparisons for \textsc{Active-VS-PE}, \textsc{Active-Bayes-PE} and \textsc{Random-PE} algorithms. Each plot represents a different configuration of the preference change scenario presented in Section~\ref{sec:pref_change}.}
    \label{fig:pref_change_combined_distance}
\end{figure*}

The results with respect to Normalized L2-distance for scenarios \textbf{Downscale-ordered}, \textbf{Upscale-ordered}, and \textbf{Random-switch} are presented in Figure~\ref{fig:pref_change_combined_distance}.
Overall, the trends are similar as the ones for accuracy metric. 

\paragraph{Additional preference change scenarios.}

Beyond the ones presented in Section~\ref{sec:pref_change}, we simulate the following other kinds of preference changes as well.
\begin{itemize}
    \item \textbf{Downscale-random}: Agent change preference to use just one random feature after timestep $t_{\text{change}}$.
    \item \textbf{Upscale-random}: Agent changes preference from using just one feature to using all features after timestep $t_{\text{change}}$.
    \item \textbf{Downscale-ordered-2}: Agent simplifies preference to use the top-two highest weighted features after timestep $t_{\text{change}}$.
    \item \textbf{Downscale-ordered-4}: Agent simplifies preference to use the top-four highest weighted features after timestep $t_{\text{change}}$.
    \item \textbf{Upscale-ordered-2}: Agent changes preference from using only two features to using all features after timestep $t_{\text{change}}$, with all features in $w^{\text{post}}$ having lower relative weight than the non-zero weights in $w^{\text{pre}}$. 
    \item \textbf{Upscale-ordered-4}: Agent changes preference from using only four features to using all features after timestep $t_{\text{change}}$, with all features in $w^{\text{post}}$ having lower relative weight than the non-zero weights in $w^{\text{pre}}$. 
\end{itemize}

\begin{figure*}[t]
    \centering
    \includegraphics[width=\linewidth]{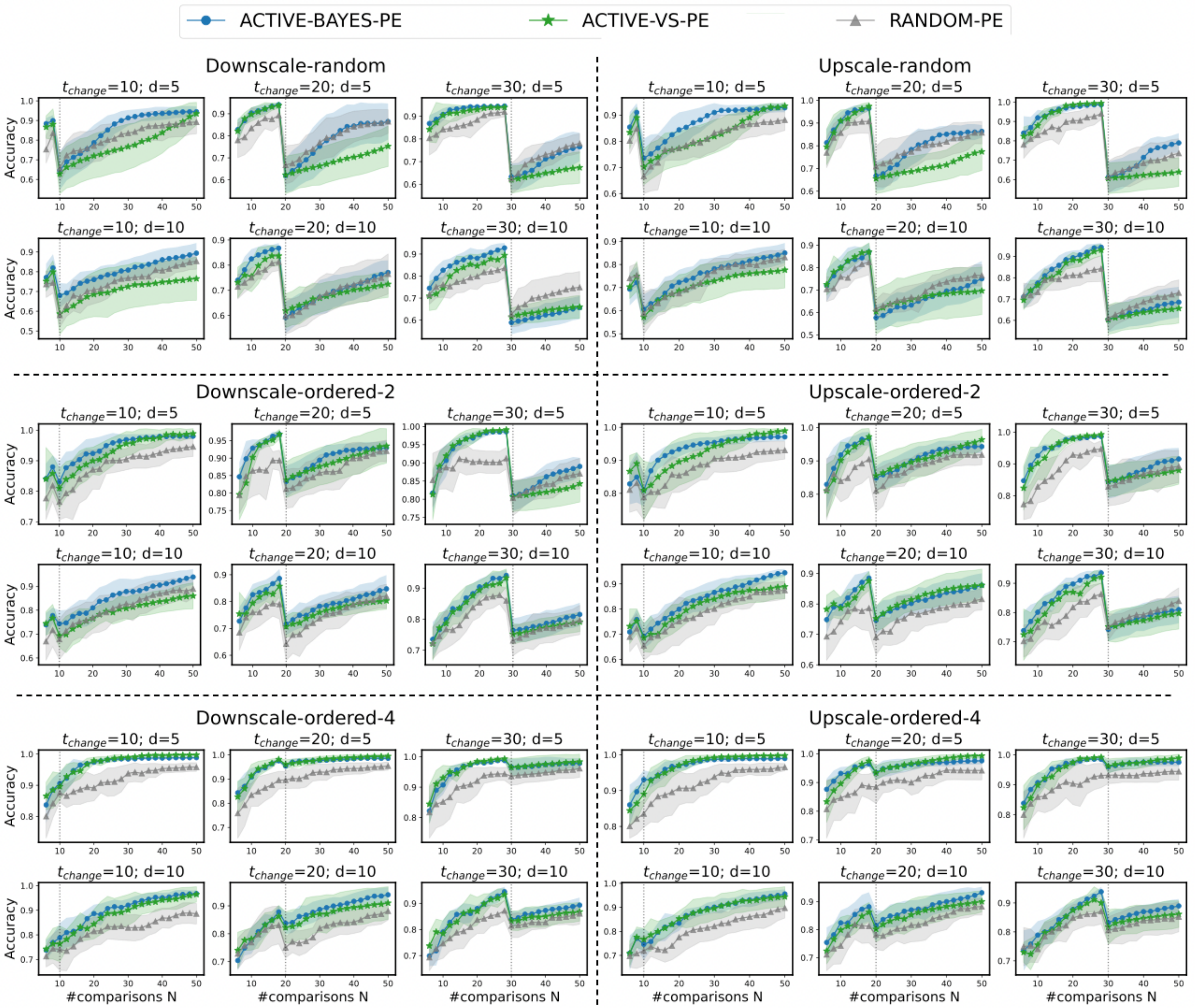}
    \caption{Accuracy vs number of comparisons for additional preference change scenarios discussed in Appendix~\ref{sec:add_results_pi}.}
    \label{fig:pref_change_combined_additional}
\end{figure*}

The results for these preference change scenarios are presented in Figure~\ref{fig:pref_change_combined_additional}.
The scenarios considered here also reflect different scales of preference change. For instance, preference change in the case of \textbf{Downscale-ordered-4} is relatively lower scale than that in \textbf{Downscale-ordered-2}. 

This preference change scale is also reflected in the results.
The accuracy drop post-$t_{\text{change}}$ is smaller in the case of \textbf{Downscale-ordered-4} and \textbf{Upscale-ordered-4}, compared to \textbf{Downscale-ordered-2} and \textbf{Upscale-ordered-2}.
Correspondingly, in these settings, the impact on active learning algorithms is also relatively smaller.
For all settings in \textbf{Downscale-ordered-4} and \textbf{Upscale-ordered-4}, one can see that both \textsc{Active-VS-PE} and \textsc{Active-Bayes-PE} recover well from preference change and have better accuracy and efficiency than the \textsc{Random-PE} baseline {\text{change}}.

As the scale of preference change increases, active learning can be seen to be less effective.
For example, the difference in accuracy between the active learning approaches and the random baseline is smaller in the case of \textbf{Downscale-ordered-2} and \textbf{Upscale-ordered-2}.
Nevertheless, we once again observe that when the number of features is small and/or preference change happens early, active learning (especially \textsc{Active-Bayes-PE}) can still achieve high accuracy faster than the random query baseline post-$t_{\text{change}}$.

% Finally, even in the case of simplification, 

\subsection{Model Misspecification} \label{sec:add_results_mm}
\begin{figure*}[!hbtp]
    \centering
    \includegraphics[width=\linewidth]{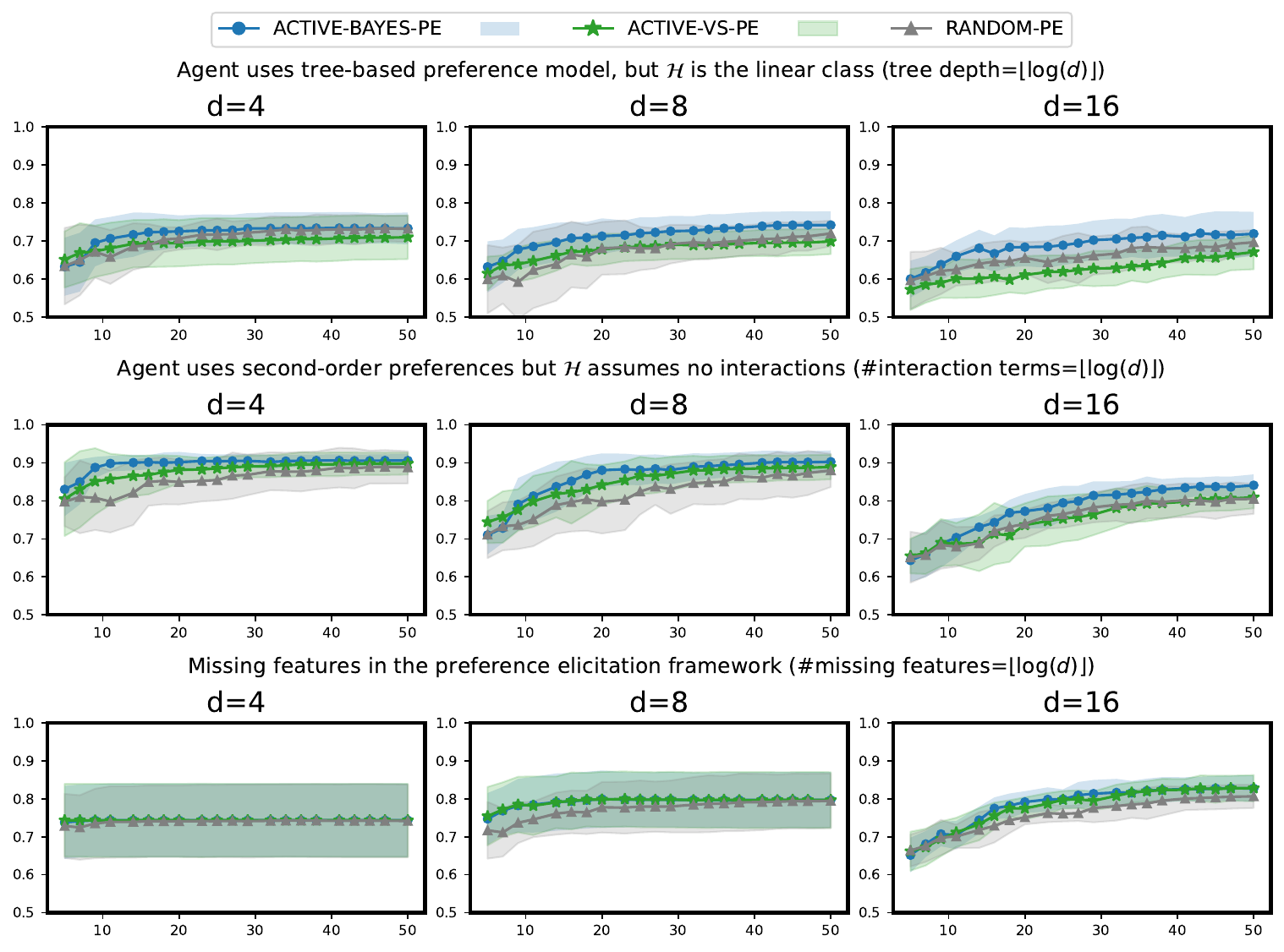}
    \caption{Accuracy vs number of comparisons for model misspecification scenarios presented in Section~\ref{sec:model_misspecify}.}
    \label{fig:model_misspecify_comb_additional}
\end{figure*}

For model misspecification scenarios, here we present results for accuracy vs number of comparisons.
We set the number of features $d$ to be 4, 8, or 16. In all cases, the extent of model misspecification is quantified by $\lfloor \log(d) \rfloor$.
The results for this setting are presented in Figure~\ref{fig:model_misspecify_comb_additional}.

\subsection{Noisy Responses} \label{sec:add_results_noise}
This section presents additional results for the simulations where the agent's responses are noisy or stochastic.

\begin{figure*}[!hbtp]
    \centering
    \includegraphics[width=\linewidth]{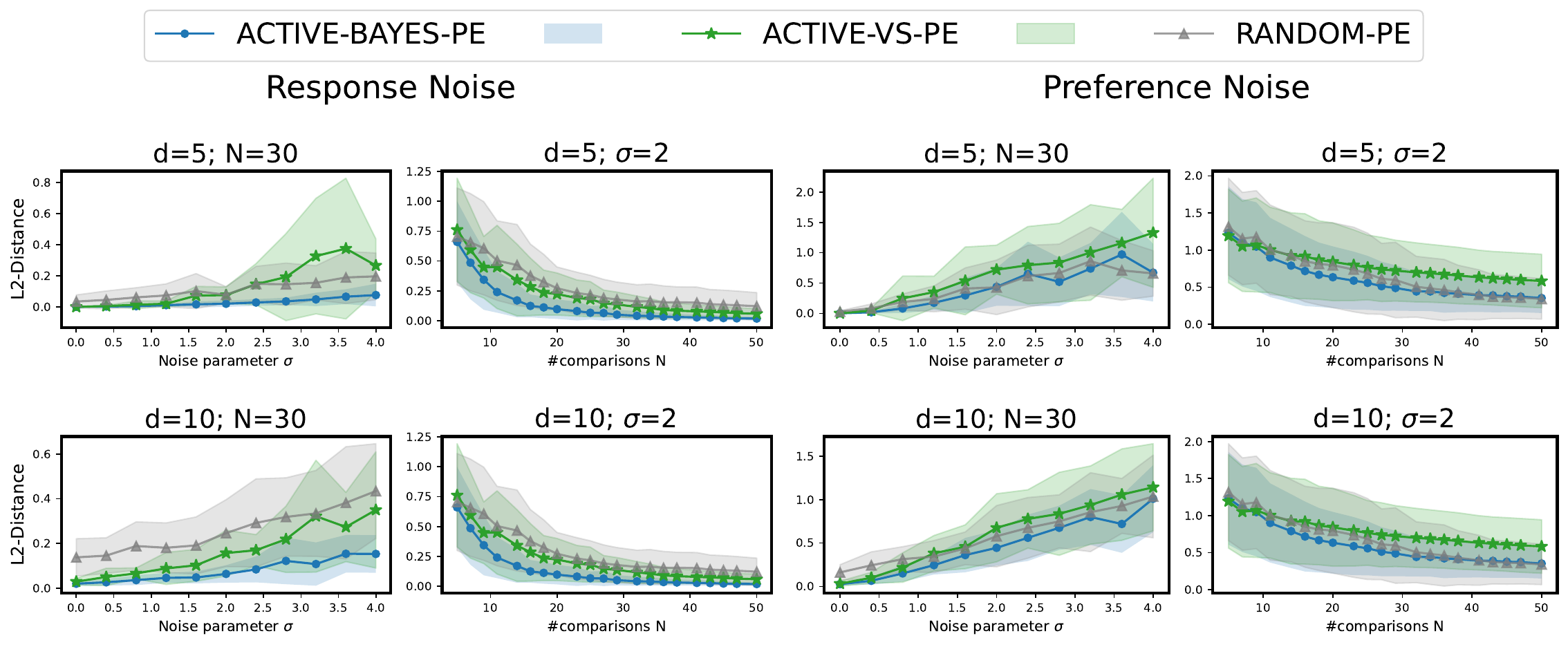}
    \caption{L2-distance comparison of \textsc{Active-VS-PE}, \textsc{Active-Bayes-PE} and \textsc{Random-PE} algorithms for different kinds of noise models presented in Section~\ref{sec:noise_analysis}.}
    \label{fig:noise_results_combined_distance}
\end{figure*}

\paragraph{Performance with respect to L2-distance.}
Figure~\ref{fig:noise_results_combined_distance} presents the equivalent of Figure~\ref{fig:noise_results_combined} for normalized distance comparison.
Here, the trends are similar to the accuracy plots. \textsc{Active-Bayes-PE} has the best performance in the case of response noise but fails to provide similarly improved performance in all settings of preference noise.

\paragraph{Performance with respect to time-variant noise.}
In addition to constant noise, we simulate the setting where the noise decreases with time.
This model can also be considered a combination of the \textit{noisy responses} and \textit{preference instability} simulations since the variation in the agent's preference/utility computation model is time-dependent.

In this case, for the \textbf{Response noise} model, $\varepsilon \sim \mathcal{N}(0, \hat{\sigma}^2)$, where $\hat{\sigma} = \sigma/\sqrt{t}$, where $t$ is the time-step/number of comparisons made so far.
Similarly, for the \textbf{Preference noise} model, $w \sim \mathcal{N}(w^\star, \hat{\sigma}^2\textbf{I}/d)$, where $\hat{\sigma} = \sigma/\sqrt{t}$.
This time we vary $\sigma$ from 1 to 10, to capture a larger range of noise parameters.

\begin{figure*}[t]
    \centering
    \includegraphics[width=\linewidth]{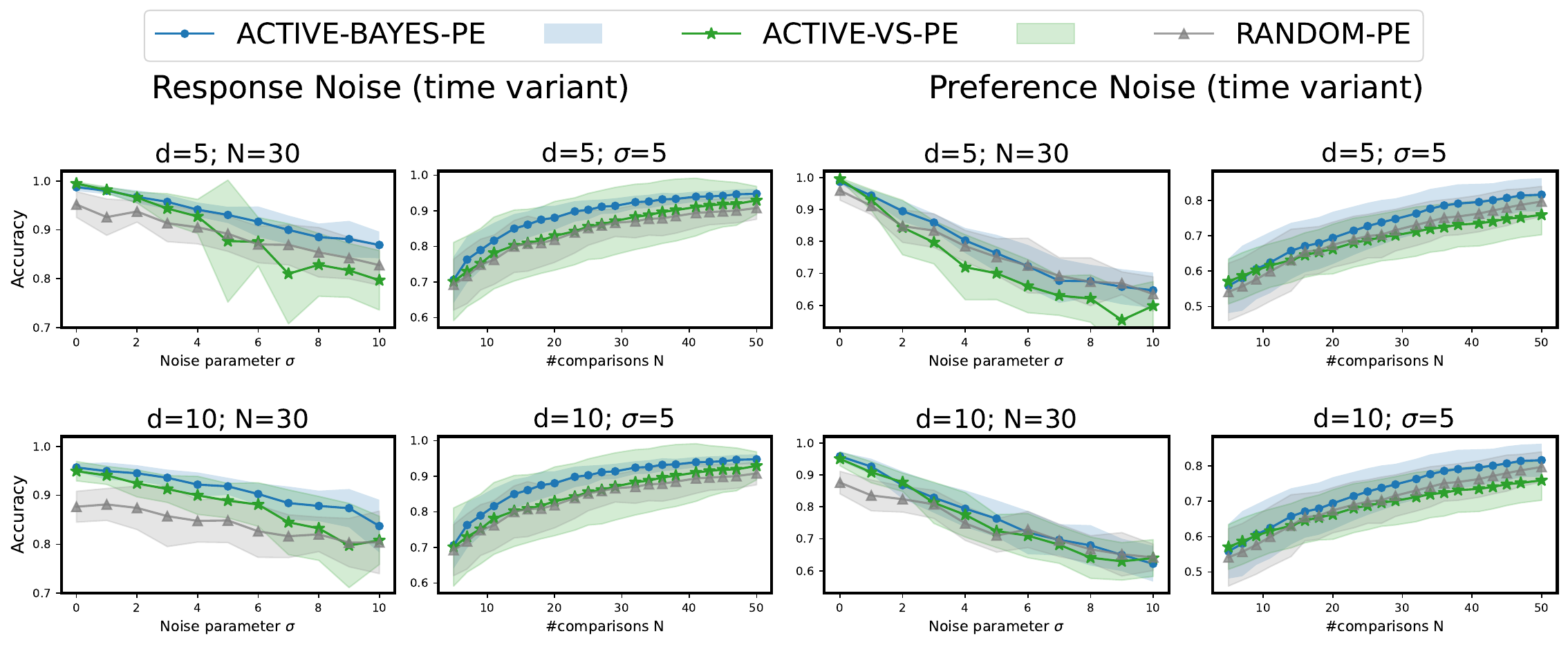}
    \caption{Accuracy of \textsc{Active-VS-PE}, \textsc{Active-Bayes-PE} and \textsc{Random-PE} algorithms for time-variant noise models.}
    \label{fig:noise_results_timed_combined}
\end{figure*}

The results for these two noise models with time-variant noise parameters are presented in Figure~\ref{fig:noise_results_timed_combined}.
Considering the relatively smaller impact of noise in this setting, the active learning algorithms have much-improved performance in comparison to the random query baseline for the response noise model.
For the preference noise model, however, improved performance of active learning is again only observed when the noise magnitude is relatively small.

\end{document}